\documentclass[
prl,
twocolumn,
superscriptaddress,
amsmath,
amssymb,
floatfix
]{revtex4-2}
\usepackage{graphicx}
\usepackage{bm}
\usepackage{physics}
\usepackage{mathtools}
\usepackage{xcolor}

\begin{document}

\title{Quantum Fisher Information as the Speed Limit for Multipartite Entanglement}

\author{Zain H. Saleem}
\affiliation{Mathematics and Computer Science Division, Argonne National Laboratory, Lemont, Illinois 60439, USA}

\author{Da-Wei Luo}
\affiliation{Center for Quantum Science and Engineering, Stevens Institute of Technology, Hoboken, New Jersey 07030, USA}

\author{Anjala M. Babu}
\affiliation{School of Physics and Applied Physics, Southern Illinois University, Carbondale, Illinois 62901, USA}

\author{Ting Yu}
\affiliation{Center for Quantum Science and Engineering, Stevens Institute of Technology, Hoboken, New Jersey 07030, USA}

\author{Stephen K. Gray}
\affiliation{Center for Nanoscale Materials, Argonne National Laboratory, Lemont, Illinois 60439, USA}

\author{Anil Shaji}
\affiliation{Indian Institute of Science Education and Research Thiruvananthapuram, Kerala 695551, India}

\begin{abstract}
We derive a universal upper bound on the speed of multipartite
entanglement generation. For arbitrary differentiable pure multipartite
states undergoing parameter-dependent evolution, we prove that the rate
of change of generalized concurrence is bounded by the square root of
the quantum Fisher information (QFI). The proof follows directly from
the symmetric logarithmic derivative formalism, revealing the bound as
a consequence of quantum estimation theory, while a complementary
Schmidt-spectrum analysis identifies the conditions for saturation and
yields a tighter bound for rank-two bipartitions. We illustrate these
results in interacting many-body spin models and investigate their
behavior across the quantum phase transition of the transverse-field
Ising model. At the critical point, we find that the QFI can be strongly
enhanced while the generalized concurrence is nearly stationary and its
sensitivity to the coupling is suppressed. This demonstrates that the
enhanced parameter sensitivity associated with quantum criticality need
not arise from rapid changes in the magnitude of multipartite
entanglement.
\end{abstract}

\maketitle

\textit{Introduction.---}
Quantum Fisher information (QFI) occupies a central role in quantum information science. It determines the ultimate precision attainable in parameter estimation through the quantum Cram\'er--Rao bound \cite{Helstrom1969,Holevo1982,Braunstein1994,Paris2009}, characterizes statistical distinguishability between neighboring quantum states, and it is proportional to the infinitesimal Bures metric on the space of quantum states except in near the points where the rank of the density matrices change \cite{Bures1969,Uhlmann1976,Braunstein1994,Hubner1992,Safranek2017,SevesoAlbarelliGenoniParis2019}. When the parameter coincides with the physical evolution time, or enters as an overall scaling of a fixed generator, QFI additionally quantifies the speed at which a quantum state traverses Hilbert space under parameter-dependent evolution, connecting quantum metrology with quantum speed limits and information geometry \cite{Mandelstam1945, Anandan1990, Margolus1998, Giovannetti2004, Giovannetti2011, DeffnerCampbell2017}.

Entanglement constitutes another cornerstone of quantum information science. Besides enabling quantum communication, computation, and simulation, multipartite entanglement is the essential resource underlying quantum-enhanced metrology \cite{Hyllus2012,Toth2012,Toth2014,Pezze2018}. While much of the existing literature has focused on how multipartite entanglement in the initial state of a quantum probe enhances the QFI in a quantum metrology scheme, increasing attention has recently turned to the dynamics of quantum resource generation including the dynamical generation of entanglement by multi-partite Hamiltonians. Universal bounds on the speed of bipartite entanglement generation have been derived using geometric and operator inequalities \cite{Mohan2023,Mohan2024}, while the growth of multipartite QFI under many-body dynamics has been shown to obey fundamental speed limits imposed by locality through Lieb--Robinson bounds \cite{Chu2023,Chu2024}. Despite these advances, a direct information-geometric relation between the instantaneous speed of multipartite entanglement generation and the QFI itself has remained unknown.

Recently, we established such a connection for arbitrary two-qubit systems. We proved that the speed of concurrence is universally bounded by the QFI and identified the corresponding saturation conditions \cite{Saleem2026Speed}. In companion works, we further related the curvature of concurrence to information-geometric robustness against parameter fluctuations, demonstrating that the QFI also bounds the loss of entanglement under weak parameter uncertainty \cite{Saleem2026Robustness,saleem2026quantum}. These results naturally raise the question of whether the same information-geometric principles persist in genuinely multipartite systems, where entanglement possesses a much richer structure and no unique scalar measure exists \cite{Coffman2000, Meyer2002, Mintert2005, Borras2006}.

In this Letter, we answer this question affirmatively. For the generalized concurrence $C_N$ defined below, we prove the universal bound
\begin{equation}
\left|\frac{dC_N}{d g}\right|\le \sqrt{F_Q}.
\label{eq:main}
\end{equation}
Our primary proof follows directly from the definition of the symmetric logarithmic derivative (SLD) operator~\cite{Helstrom1967, Helstrom1969, Holevo1973, Holevo1982, Braunstein1994}, revealing the bound as a natural consequence of quantum estimation theory. A complementary derivation based on the Schmidt decomposition \cite{Schmidt1907, EkertKnight1995} provides additional physical insight, identifies the equality conditions, and clarifies the behavior near separable states. We emphasize here that our proof of Eq.~\eqref{eq:main} is independent of the manner in which the parameter $g$ enters the Hamiltonian of the quantum system of interest.

\textit{Proof.---} We start from the symmetric logarithmic derivative (SLD) operator $L$, defined by
\begin{equation}
\partial_g\rho = \frac12(L\rho+\rho L),
\label{eq:sld}
\end{equation}
in terms which we can write the QFI as
\begin{equation}
F_Q = {\rm Tr}(\rho L^2) = \langle L^2 \rangle.
\label{eq:qfisld}
\end{equation}
The eigenbasis of $L$ defines the locally optimal measurement saturating the quantum Cram\'er--Rao bound. 

To quantify multipartite entanglement, we employ the generalized concurrence introduced in Refs.~\cite{Mintert2005,Borras2006}, which is constructed from the linear entropies of all inequivalent bipartitions. For a bipartition $i$, the concurrence is
\begin{equation}
C_i=\sqrt{2\left(1-\mathrm{Tr}\rho_i^2\right)},
\label{eq:Ci}
\end{equation}
where $\rho_i$ denotes either reduced density matrix associated with that bipartition. The multipartite concurrence is then defined as
\begin{equation}
C_N = \sqrt{\frac{1}{M}\sum_i C_i^2} = \sqrt{\frac{2}{M}\sum_i \left(1-\mathrm{Tr}\rho_i^2\right)},
\label{eq:CN}
\end{equation}
where $M=2^{N-1}-1$ is the number of in-equivalent bipartitions. This normalization averages over all distinct bipartitions, avoids double counting complementary partitions, and reduces to the standard two-qubit concurrence when $N=2$.

We first prove the bound for an arbitrary bipartition $A|B$. Let $\rho_A={\rm Tr}_B\rho$ and
\begin{equation}
C^2=2\left(1-\mathrm{Tr}\rho_A^2\right).
\label{eq:C2}
\end{equation}
At points where $C>0$, differentiation gives
\begin{equation}
\partial_g C=-\frac{2}{C}\mathrm{Tr}(\rho_A \partial_g \rho_A)
=-\frac{2}{C}\mathrm{Tr}\!\big[(\rho_A\otimes \openone_B) \partial_g \rho\big].
\label{eq:CdotTrace}
\end{equation}
Substituting the SLD relation \eqref{eq:sld} into Eq.~\eqref{eq:CdotTrace}, and defining $A=\rho_A\otimes \openone_B$, yields
\begin{equation}
\label{eq:ineq1a}
\partial_g C=-\frac{2}{C}\mathrm{Re}\langle A L\rangle \leq -\frac{2}{C} \langle AL \rangle.
\end{equation}
Since $\langle L\rangle={\rm Tr}(\rho L)=0$, we may write $\langle A L\rangle=\langle(A-\langle A\rangle)L\rangle$. The Cauchy--Schwarz inequality then gives
\begin{equation}
|\partial_g C|\le \frac{2\Delta A}{C}\sqrt{\langle L^2\rangle}
=\frac{2\Delta A}{C}\sqrt{F_Q}.
\label{eq:almost}
\end{equation}
It remains to bound the pre factor.

Let $\{\lambda_i\}$ be the eigenvalues of $\rho_A$. Then
\begin{equation}
(\Delta A)^2=\sum_i\lambda_i^3-\Big(\sum_i\lambda_i^2\Big)^2
=\sum_{i<j}\lambda_i\lambda_j(\lambda_i-\lambda_j)^2.
\label{eq:varianceidentity}
\end{equation}
Since $(\lambda_i-\lambda_j)^2\le 1$,
\begin{equation}
\label{eq:ineq3a}
(\Delta A)^2\le \sum_{i<j}\lambda_i\lambda_j=\frac{C^2}{4}.
\end{equation}
Substitution into Eq.~\eqref{eq:almost} proves
\begin{equation}
|\partial_g C_i|\le \sqrt{F_Q}
\label{eq:bipbound}
\end{equation}
for every bipartition $i$.

The multipartite result follows from Eq.~\eqref{eq:CN}. Since
\begin{equation}
\partial_g C_N=\frac{1}{M C_N}\sum_i C_i \, \partial_g C_i,
\end{equation}
we have, using Eq.~\eqref{eq:bipbound} and the Cauchy--Schwarz inequality,
\begin{equation}
\label{eq:ineq4a}
|\partial_g C_N| \le \frac{\sqrt{F_Q}}{M C_N}\sum_i C_i \le \sqrt{F_Q}.
\end{equation}
This proves Eq.~\eqref{eq:main}. \hfill$\square$

\textit{Saturation conditions.---}
The proof of $|\partial_g C_N| \leq \sqrt{F_Q}$ uses a sequence of four inequalities, all of which must be saturated if $|\partial_g C_N|$ is to be equal to $\sqrt{F_Q}$. The first inequality that is used in Eq.~\eqref{eq:ineq1a}  is $| \mathrm{Re} \langle{A_iL}\rangle_\rho | \le  | \langle {A_iL} \rangle_\rho|$ for every bipartition and therefore, for saturation we require   $\mathrm{Im}\,\Tr(\rho A_iL)=0$. Since $\Tr(\rho A_iL)^*=\Tr(\rho LA_i)$, this is equivalent to $\Tr\big(\rho\,[A_i,L]\big)=0$. We call this saturation condition (ii). The second inequality that appears in Eq.~\eqref{eq:almost}, is the Cauchy-Schwarz one, $|\langle {(A_i-\langle {A_i} \rangle)L } \rangle_\rho | \le \Delta A_i\,\sqrt{F_Q}$. Because $\rho=\ket\psi\bra\psi$ has rank one, its support is the single ray $\mathrm{span} \{ \ket\psi \}$, and equality holds iff, 
\begin{equation}
    \label{eq:ineq2}
    (A_i-\langle{A_i}\rangle )\ket\psi \;=\; c_i\,L\ket\psi \;=\; 2c\ket{\partial_g \psi},
\end{equation} 
for some real constant $c$. This is saturation condition (ii).  Note that this has to hold for every bipartition on the left side while on the right side we have a constant vector making the saturation condition rather restrictive. Saturation of \eqref{eq:ineq2}, for bipartition $A|B$, requires the local Schmidt bases on both $A$ and $B$ to be instantaneously stationary. Conditions (i) and (ii) together also imply that for saturation, the relative phases of the Schmidt coefficients are constrained as well. A detailed analysis of conditions (i) and (ii) is given in the end matter.

The third inequality appears in Eq.~\eqref{eq:ineq3a}. Every pair contributing to the sum in Eq.~\eqref{eq:ineq3a} has $\lambda_k,\lambda_l>0$. Since $\lambda_k,\lambda_l\ge0$ and $\lambda_k+\lambda_l\le1$, $|\lambda_k-\lambda_l| \le \lambda_k+\lambda_l \le 1$,
with $|\lambda_k-\lambda_l|=1$ possible only when $\min(\lambda_k \lambda_l ) = 0$; but then $\lambda_k\lambda_l=0$ and the pair does not contribute to the sum. Consequently this inequality is never exactly saturated at any point with $C_i>0$; it is only approached in the limit $C_i\to0$ when the bipartition tends to a product state. We call this saturation condition (iii). A strictly tighter, exactly-saturable bound exists whenever $\rho_{A_i}$ has Schmidt rank exactly two (e.g.\ any single-qubit-vs-rest bipartition): using the exact value $(\lambda_1-\lambda_2)^2=1-C_i^2$ in place of the bound $\le1$,
\begin{eqnarray}
	(\Delta A_i)^2 &  = &  \lambda_1\lambda_2(1-C_i^2) = \frac{C_i^2}{4}(1-C_i^2) \nonumber \\
    & \Rightarrow &  |\partial_g C_i|\le \sqrt{1-C_i^2} \, \sqrt{F_Q}.
\label{eq:tightbound}
\end{eqnarray}
This tight bound is always available for the two qubit case. Saturation condition (iv) appears in~\eqref{eq:ineq4a}, where we require $\sum_iC_i =  MC_N$. This condition is satisfied only if every bipartition carries exactly the same amount of bipartite entanglement at any instant with $C_1=C_2=\cdots=C_M$. For saturating the inequality~\eqref{eq:ineq4a} we further require that all nonzero terms $C_i \partial_g C_i$ have the same sign so that $|\sum_i C_i \partial_g C_i| = \sum_i C_i |\partial_g C_i|$. This is saturation condition (v) which implies that all the constants $c_i$ that appear in Eq.~\eqref{eq:ineq2} have the same sign.

Putting it all together, exact saturation of $|\partial_g C_N|=\sqrt{F_Q}$ requires, simultaneously, for every bipartition $i$ with $C_i>0$,  $\Tr(\rho[A_i,L])=0$, local Schmidt bases on both sides of cut $i$ instantaneously stationary, the Schmidt coefficients do not pick up relative phases, $C_i \to 0$ and all $C_i$ equal with all constants $c_i$ having the same sign. Saturation conditions (iii) and (iv) together are consistent only with $C_i\to0$ for \emph{all} bipartitions simultaneously and so we expect exact saturation of the stated bound to be confined to the product-state boundary of the dynamics. Away from that boundary, the remaining  conditions can still be met but the bound is not saturated. However, the \emph{tight}, rank-2-specific bound Eq.~\eqref{eq:tightbound} can be saturated away from the product state boundary.

\textit{Isotropic $N$-qubit Pauli interaction Hamiltonian.---}
Dynamics of the Hamitonian, 
\begin{align}
    H = g\sum_{j=x,y,z}\bigotimes{}_{i=1}^{N}\sigma_j^{(i)} = g\tilde{h}, \label{eq_hxyz}
\end{align}
is instructive since it illustrates the conditions for saturation of both the generic bound \eqref{eq:main} as well as the rank-2 specific bound, \eqref{eq:tightbound}. This Hamiltonian corresponds to the isotropic Heisenberg
spin interaction for $N = 2$, but for $N > 2$ it does not correspond to typical condensed matter spin
systems.  The types of terms in this Hamiltonian, however, do appear in stabilizer generators~\cite{Nielsen2010}. The state of the system at any time, $t$ depends on $g$ and $t$ only through the combination $gt$. 

Fix a Pauli axis $j\in\{x,y,z\}$. Let the initial state be
$|\psi_0\rangle=\bigotimes_{k=1}^N |\phi^{(k)}_{j,s_k }\rangle$, where $|\phi^{(k)}_{j,s_k}\rangle$ is an eigenstate of one of the three Pauli operators, $\sigma_j$, with eigenvalue $s_k=\pm1$, which may vary from
site to site. Thus, this class includes ferromagnetic product states as well as the N\'eel state, for which $j=z$ and $s_k=(-1)^k$. Let $|\phi^{(k)}_{j,-s_k}\rangle$ be the orthogonal eigenstate of $\sigma_j$ with eigenvalue $-s_k$. We denote $\bigotimes_{k=1}^N |\phi^{(k)}_{j,-s_k }\rangle = |\psi_0^\perp\rangle$. Action of the Pauli operators on any $|\phi^{(k)}_{j,s_k} \rangle$ either leaves it invariant or maps it to $|\phi^{(k)}_{j,-s_k}\rangle$ up to a phase. Hence,
$\tilde{h}\bigotimes_{k=1}^N|\phi^{(k)}_{j,s_k}\rangle$ and
$\tilde{h}\bigotimes_{k=1}^N|\phi^{(k)}_{j,-s_k}\rangle$ remain within $\mathrm{span}{|\psi_0\rangle, |\psi_0^\perp \rangle}$.  This means that the $2^N$-dimensional problem collapses onto a two-level  problem for all time, for any such initial state $|\psi_0\rangle$.

Since $|\psi_0\rangle$ and $|\psi_0^\perp \rangle$ are both product states differing at \emph{every} site, their restrictions to any nonempty subset $A$ are automatically orthogonal for every bipartition $A|B$. Hence $\ket{\psi(\theta)} = \alpha(\theta) |\psi_0 \rangle + \beta(\theta) |\psi_0^\perp \rangle $ is \emph{already} in Schmidt form, with local bases $\{
\bigotimes_{k\in A}| \phi_{j,s_k}^{(k)} \rangle,\, \bigotimes_{k\in A} |\phi_{j,-s_k}^{(k)} \rangle \}$ and
$\{ \bigotimes_{k\in B}|\phi_{j,s_k}^{(k)} \rangle, \, \bigotimes_{k\in B} |\phi_{j,-s_k}^{(k)}\rangle \}$
that never rotate, for every one of the $M=2^{N-1}-1$
bipartitions at once and for all $\theta$. For satisfying saturation conditions (i) and (ii) in this effective two dimensional case, we further require that $\alpha(\theta)$ and $\beta(\theta)$ does not pick up a $\theta$ dependent relative phase.  Every $\rho_{A_i}$ has Schmidt rank exactly two, with eigenvalues $\{|\alpha|^2,|\beta|^2\}\equiv\{1-P,P\}$ identical across every bipartition. Therefore, condition that all $C_i$ are equal holds exactly. Note that, if all the saturation conditions except (iii) hold, then the \emph{tight} bound Eq.~\eqref{eq:tightbound} applies with no loss, globally since the dynamics is confined to a two dimensional subspace of the entire Hilbert space. The general bound $|\partial_g C_N|\leq\sqrt{F_Q}$ is therefore touched exactly only at the isolated instants where $C_N=0$.

\begin{figure}[!htb]
    \centering
    \includegraphics[width=.45\textwidth]{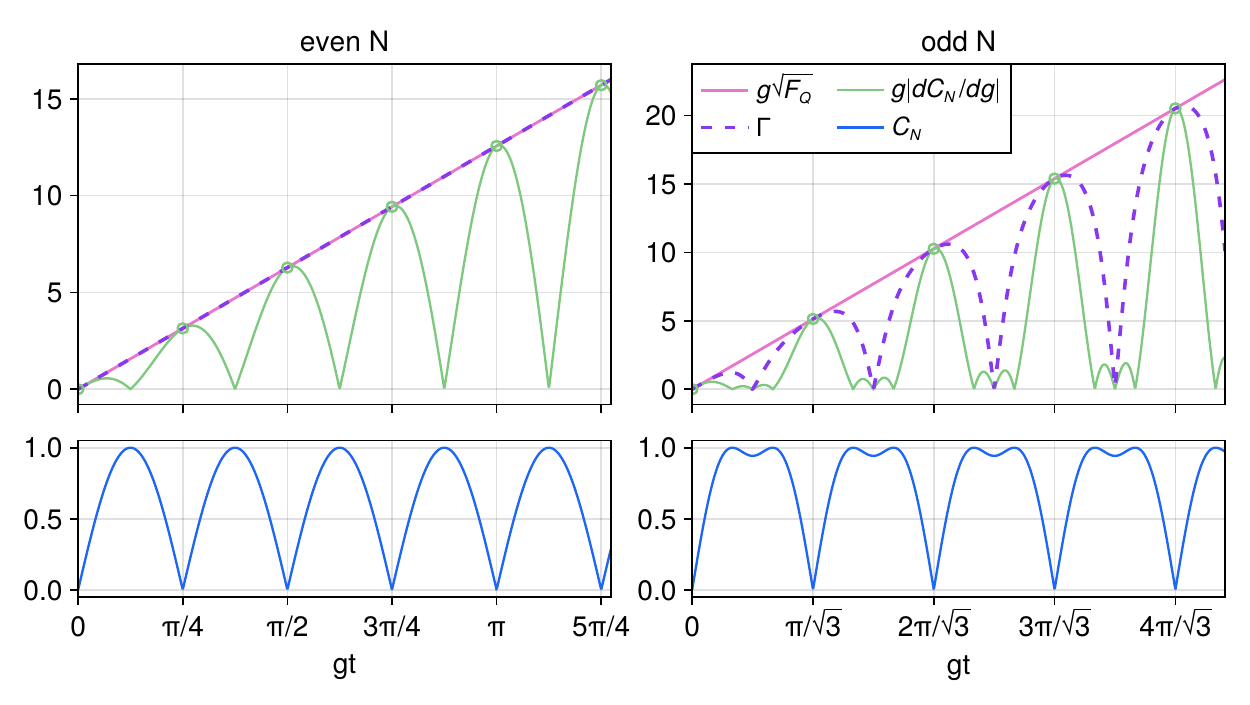}
    \caption{Dynamics of multipartite entanglement, its derivative, and QFI  under the Hamiltonian Eq.~\eqref{eq_hxyz} with an initial N\'{e}el state of alternating up and down spins. Different behaviors are observed according to the \textit{parity} of \(N\), and in both cases, the square root of the QFI serves as the envelope for $|dC_N/dg|$, where \(|dC_N/dg| \rightarrow \sqrt{F_Q}\) when $C \rightarrow 0$. Furthermore as discussed in the text, the dynamics of the QFI and concurrence are the same for all $N$ within a given parity. Blank circles indicates points where the derivative does not exist. The tighter bound Eq.~\eqref{eq:tightbound} is also shown as the purple dashed line, where we define \(\Gamma = |dC_N/dg| / \sqrt{1-C_N^2} \). In the lower panels, the dynamics of the entanglement is plotted as a function of time.\label{fig_dcdg}}
\end{figure}

In Fig.~\ref{fig_dcdg} we show the dynamics of multipartite concurrence $C$, its first-order derivative, and the quantum Fisher information for the isotropic $N$-qubit Pauli interaction Hamiltonian with the initial N\'{e}el state $|\psi(0)\rangle=\bigotimes_{k=1}^N |k \bmod 2\rangle$, i.e., a state with alternating up and down spins. This is a special case of the states of the form
$\bigotimes_{k=1}^N |\phi_{j,s_k}^{(k)}\rangle$ considered above, with $j=z$ and $s_k=(-1)^k$, such that all $|\phi_{j,s_k}^{(k)} \rangle$ are eigenstates of the same Pauli operator $\sigma_z$. For the N\'{e}el state, we see that $N$-qubit systems with even and odd numbers of qubits display different behavior. However, in both cases the first-order derivative of concurrence lies under the
envelope of $\sqrt{F_Q}$, and $|dC/dg| \rightarrow \sqrt{F_Q}$ when $C\rightarrow0$. Another interesting fact is that all even (odd) $N$ gives the same values of \(F_Q\) and \(C\): for this particular choice of Hamiltonian and initial state, the QFI and entanglement dynamics are only dependent on the parity of $N$ and are the same for all even (odd) $N$. The key difference between the even and odd cases is that for the odd case, the Schmidt coefficients picks up a relative phase ($\alpha_1 + \beta_1 \neq \alpha_2 + \beta_2$) while in the even case they do not. Because of this the tight bound is globally satisfied for the even case while the same is not true for the odd case as can be seen from Fig.~\ref{fig_dcdg}.

Note that with a Hamiltonian that is proportionate to $g$ as in this and the next example, the unit-less quantities $\sqrt{F_Q}~g $ and $|dC_N/dg|~g$ when plotted as functions of $gt$ are universal, and from a numerical point of view a calculation with a particular $g$, say $g$ = 1, suffices to provide information about all $g$ from such plots. (See also discussion in Appendix A of \cite{Saleem2024Achieving}.)

From Fig.~\ref{fig_dcdg} we also see that the square root of the QFI is a monotonically increasing function of $gt$ and so are the periodic maxima of $|\partial_g C_N|$. However, the Hilbert space of the quantum probe is finite dimensional and neither the QFI nor the rate of generation of entanglement can grow in an unbounded manner. The monotonic increase in the QFI, which is a measure of distinguishability of the time evolved probe states corresponding to $g$ and $g'= g + \delta g$ should be understood as in increase in the ability to distinguish between $g$ and $g'$ separated by smaller values of $\delta g$ (increasing resolution in $g$) as $t$ increases. In the same way the increase in the peak values of $|\partial_g C_N|$ indicate that time evolved states of the probe corresponding to $g$ and $g'$ can differ substantially in their entanglement content as $t$ increases.

\begin{figure}[!htb]
    \centering
    \includegraphics[width=0.95\linewidth]{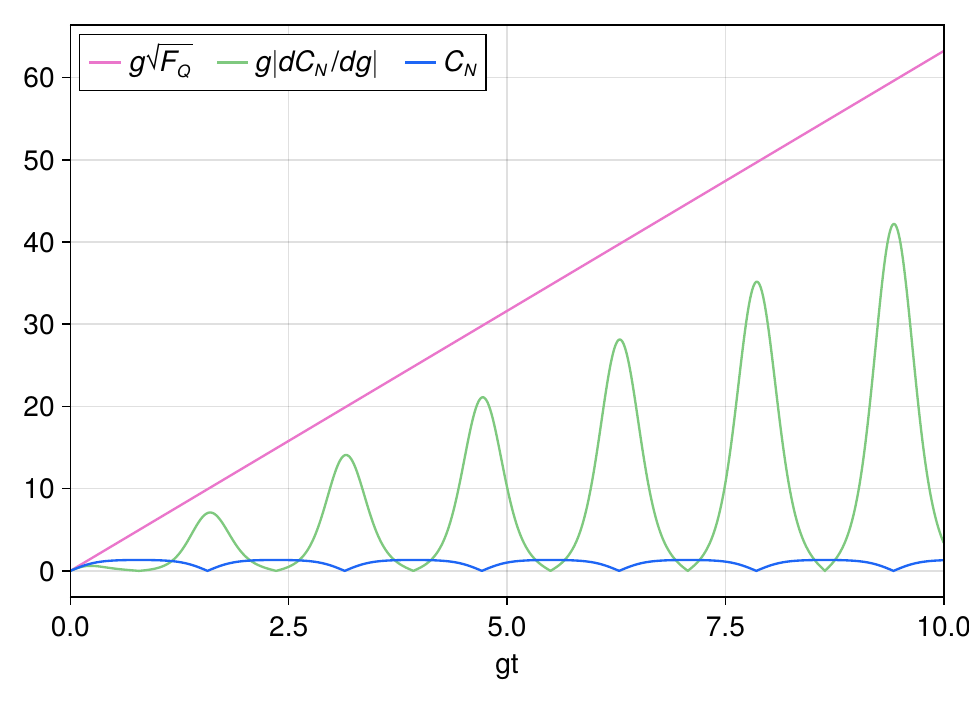}
    \caption{The concurrence $C_N$, $|\partial_g C_N|$ and the $\sqrt{F_Q}$ is plotted for a chain of $N$ = 10 qubits with periodic boundary conditions evolving from the initial state $|+\rangle^{\otimes N}$ under the Ising Hamiltonian.}
    \label{fig:Isingfigs}
\end{figure}

{\textit{The Ising model.---}}  We now investigate entanglement generation in a model commonly used in solid-state physics. The Ising model Hamiltonian for $N$ qubits/spins is, 
\begin{equation}
H(g) = -g\sum_{i=1}^{N}\sigma^z_i\sigma^z_{i+1} = g\tilde{h}.
\label{eq:Ising}
\end{equation}
In Fig.~\ref{fig:Isingfigs} we have plotted the concurrence,  $|\partial_g C_N|$ and the square root of the QFI for a chain of $N$ = 10 qubits with periodic boundary conditions, initialized in the state $|+ \rangle^{\otimes N}.$ We see that the bound is not saturated in this case but is still useful as the maxima of $|\partial_g C_N|$ do track the bound reasonably well. 

{\textit{Transverse-field Ising model.---}} 
Investigating the bound \eqref{eq:main} for the transverse-field Ising model (TFIM) is instructive since it throws light on the question of how multipartite entanglement behaves around a critical point. On a chain of $N$ qubits/spins, the TFIM Hamiltonian is
\begin{equation}
H(g,h) = -g\sum_{i=1}^{N}\sigma^z_i\sigma^z_{i+1} - h\sum_{i=1}^N\sigma^x_i,
\label{eq:HTFIM}
\end{equation}
with $g$ the nearest-neighbor Ising coupling and $h$ the transverse field. We assume periodic boundary conditions for the chain for our numerical investigations. The first term favors ferromagnetic alignment along $z$ while the second is an external competing field along the transverse ($x$) direction that does not commute with the first. The transverse field drives quantum fluctuations that can destroy the ferromagnetic order even at zero temperature. The model is of interest both as an idealization of real quasi-one-dimensional magnetic materials and, more importantly for us, as a minimal setting for a continuous quantum phase transition which is driven by the competition between two non-commuting terms in the Hamiltonian rather than by thermal fluctuations \cite{Sachdev2011}. The model is exactly solvable using a  Jordan--Wigner mapping of spins to spinless fermions \cite{Pfeuty1970, JordanWigner1928, LiebSchultzMattis1961}.

\begin{figure}[!htb]
    \centering
    \includegraphics[width=0.86\linewidth]{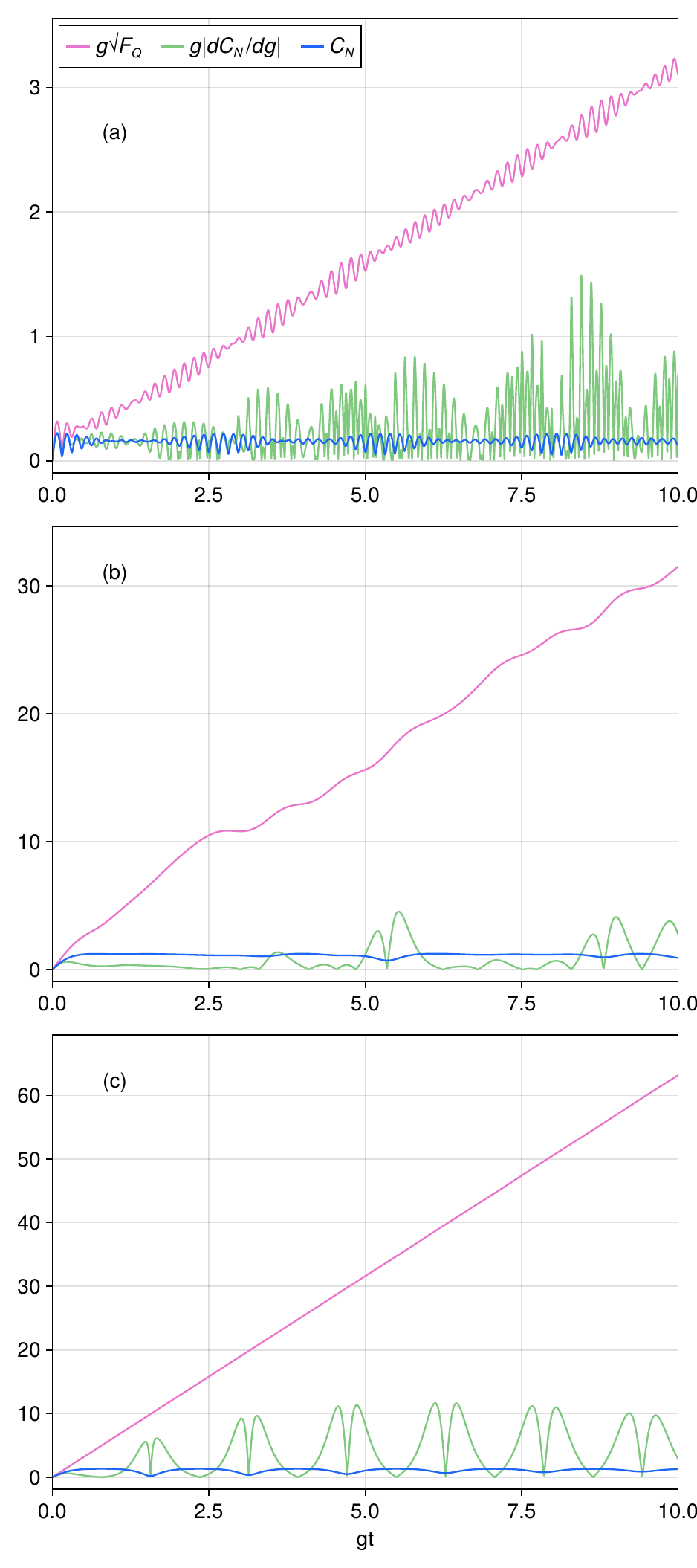}
    \caption{The concurrence $C_N$, $|\partial_g C_N|$ and the $\sqrt{F_Q}$ is plotted for a chain of 10 qubits with periodic boundary conditions evolving from the initial state $|+\rangle^{\otimes N}$ under the TFIM Hamiltonian is plotted for $g=0.1$ (a), $g=1$ which corresponds to the critical point (b) and $g=20$ (c). On the $x$ axis time has been scaled to $gt$. We see that away from the the critical point $|\partial_g C_N|$ does change and has values that come closer to $\sqrt{F_Q}$ compared to the case where $g=1$. We see that in all cases the bound is quite far from being saturated with the bound being particularly loose at the critical point. We also see the transition from the regime dominated by the transverse field in panel (a) to the one dominated by the nearest neighbor coupling in panel (c). In both these cases, the two energy and time scales that are involved creates oscillations with beat notes while at the critical point when the two scales are identical the behavior of concurrence and its derivative are non-oscillatory.}
    \label{fig:TFIMfigs}
\end{figure}

The quantum phase transition from the ferromagnetically ordered phase for $g>h$, with $\langle \sigma^z_i \sigma^z_j \rangle $ approaching a nonzero constant at large separation and a paramagnetic phase for $g<h$, with the field-aligned product state as the $g \to 0$ limit occurs at $g=h$. In Fig.~\ref{fig:TFIMfigs} we have plotted the concurrence,  $|\partial_g C_N|$ and the square root of the QFI for three $N$ = 10 cases corresponding to $g$ having values 0.1, 1 and 20 respectively with $h=1$. We see that for values of $g$ far from the critical point at $g=h=1$, the concurrence as well as $|\partial_g C_N|$ are both varying as a function of time while at the critical point, the concurrence itself is almost stationary and $|\partial_g C_N|$ is close to zero. Furthermore, the difference between $\sqrt{F_Q}$ and $|\partial_g C_N|$ is greater at the critical point compared to the other cases, indicating that at the critical point, the sensitivity of entanglement to changes in $g$ is quite low. This means that, contrary to what may be naively expected, significant changes in the magnitude of $N$-party entanglement (dynamically generated entanglement) may not be the reason for the sensitivity of a quantum probe around the critical point to small changes in $g$~\cite{ZanardiPaunkovic2006, ZanardiGiordaCozzini2007, ZanardiParisVenuti2008, InvernizziKorbiczParis2008, FrerotRoscilde2018, RamsSierantDuttaHorodeckiZakrzewski2018, GietkaMetzKellerLi2021, GarbeBinaKellerParisFelicetti2020, HaukeHeylTagliacozzoZoller2016, GuanLewisSwan2021, ZhouKongLanZhang2023, IliasYangHuelgaPlenio2022, LiuChenJiangEtAl2021, DingLiuShiGuoMolmerAdams2022, GhoshKonarRakshitSenDeSen2026, MihailescuAlushiDiCandiaFelicettiGietka2025, MontenegroMukhopadhyayYousefjaniSarkarMishraParisBayat2024}. 

The TFIM Hamiltonian is not of the form $g\tilde h$ and it contains two non commuting terms, only one of which is proportional to $g$. Consequently $g$ and $t$ cannot be treated as interchangeable unlike in the previous example. Evolving with a coupling $g$ for a longer time, and evolving a larger coupling $g'$ for the shorter time are therefore genuinely different trajectories in Hilbert space as can be seen from the differences between the three panels in Fig.~\ref{fig:TFIMfigs}, all of which are plotted against the scaled time coordinate, $gt$. For this example, $F_Q (g,t) \neq 4t^2\,\mathrm{Var}_{\rho_{t}} (\partial_g H)$ but instead of $\partial_g H$, we have to insert into this expression,  the generator of translations in $g$ at fixed $t$ which is given by 
\[ \mathcal G(g,t) \! = \! U(g,t)\Big[ \! \int_0^t d\tau \, e^{iH(g,h)\tau} (\partial_gH) e^{-iH(g,h)\tau} \! \Big]U(g,t)^\dagger.\]

\textit{Discussion.---}
Quantum Fisher information is usually interpreted as the quantity governing the ultimate precision of quantum parameter estimation. The present work demonstrates that its significance is broader: for arbitrary differentiable pure multipartite states, QFI also bounds the rate at which multipartite entanglement may be generated.

The proof is entirely information-geometric. By expressing the derivative of generalized concurrence through the SLD, the entanglement speed bound emerges from the same mathematical structure responsible for the quantum Cram\'er--Rao bound. The Schmidt-spectrum formulation provides a complementary interpretation in terms of optimal redistribution of Schmidt coefficients.

Several extensions remain open. An important direction is the derivation of corresponding bounds for mixed states, where the Bures metric rather than the Fubini--Study metric determines the geometry of quantum evolution. It would also be interesting to determine whether analogous speed limits exist for other multipartite entanglement monotones and whether higher-order derivatives admit universal geometric bounds.

Our results identify QFI as the fundamental quantity governing not only how precisely a parameter can be estimated, but also how rapidly multipartite entanglement can be created. They therefore establish a direct bridge between quantum metrology, multipartite entanglement theory, and information geometry.

\begin{acknowledgments}
The authors acknowledge useful discussions with collaborators and colleagues on quantum metrology and multipartite entanglement. Z.H.S. acknowledges support from the U.S. Department of Energy, Office of Science, Advanced Scientific Computing Research (ASCR), under Contract No. DE-AC02-06CH11357. Work performed at the Center for Nanoscale Materials, a U.S. Department of Energy Office of Science User Facility, was supported by the U.S. DOE, Office of Basic Energy Sciences, under Contract No. DE-AC02-06CH11357.
\end{acknowledgments}

\bibliography{references}

\appendix

\section{End matter: Saturation conditions (i) and (ii) }\label{appA}
It is worth taking a closer look at the saturation conditions given in \eqref{eq:ineq1a} and \eqref{eq:ineq2}. We write $\ket\psi = \sum_k \sqrt{\lambda_k} \ket{k}_A \ket{k}_B$ in the instantaneous Schmidt basis for bipartition $A|B$. Normalization of the Schmidt basis states means that ${\rm Re}[\langle k|\partial_g k\rangle] = 0$ for both subsystems. Denoting by $|k^\perp \rangle$ a vector in the subspace perpendicular to $|k\rangle$ we write 
\begin{eqnarray*}
    |\partial_g k\rangle_A & = & i\alpha_k|k\rangle_A + |\partial_g k_A^\perp\rangle, \\
    |\partial_g k\rangle_B & = & i\beta_k|k\rangle_B + |\partial_g k_B^\perp\rangle.
\end{eqnarray*}  
Since $A$ acts trivially on $B$, using the Schmidt form of $|\psi\rangle$, we can write saturation condition (ii) as,
\begin{equation}
	\big(A-\langle{A} \rangle \big) \ket\psi = \sum_k \sqrt{\lambda_k}\,(\lambda_k - \langle A \rangle )\ket{k}_A \ket{k}_B,
\end{equation}
which lies entirely in the \emph{diagonal} subspace $\mathrm{span}\{\ket{k}_A, \ket{k}_B\}$. Expanding $\ket{\partial_g \psi}$ in general,
\begin{eqnarray}
    \label{eq:appendix1}
	\ket{\partial_g \psi} & = &  \sum_k\frac{\partial_g \lambda_k}{2\sqrt{\lambda_k}}\ket k_A\ket k_B  \nonumber  \\
    && \; + \sum_k \sqrt{\lambda_k} \big( \ket{\partial_g k}_A \ket k_B + \ket k_A \ket{\partial_g k}_B \big), \nonumber \\
    & = &  \sum_k \bigg[ \frac{\partial_g \lambda_k}{2\sqrt{\lambda_k}} + i\sqrt{\lambda_k}(\alpha_k + \beta_k)\bigg]\ket k_A\ket k_B \nonumber \\
    &&  + \sum_k \sqrt{\lambda_k} \big( \ket{\partial_g k^\perp}_A \!\ket k_B + \ket k_A \! \ket{\partial_g k^\perp}_B \!\big). \qquad
\end{eqnarray}
Using $\braket{k}{\partial_g k}=0$ (normalization), the second group of terms is automatically \emph{off-diagonal}. Equation~\eqref{eq:ineq2} therefore forces $\ket{\partial_g k^\perp}_A = 0$ and $\ket{\partial_g k^\perp}_B = 0$ at each instant. Saturation of \eqref{eq:ineq2}, for bipartition $A|B$, requires the local Schmidt bases on both $A$ and $B$ to be instantaneously stationary: $g$ may re-weight the existing Schmidt coefficients but may not rotate the local bases. However, most interesting entangling Hamiltonians do rotate the instantaneous Schmidt basis  and so this is a rather restrictive condition.

Saturation conditions (i) and (ii) puts further constraints on $\partial_g \lambda_k$, $\alpha_k$ and $\beta_k$. However, before exploring these, we note that all three quantities have a ``gauge" freedom since $|\psi\rangle$ is defined only up to an overall phase. This freedom can be removed by fixing the gauge using the condition $\langle \psi | \partial_g \psi \rangle = 0$. Imposing the gauge condition, leads to $\sum_k \partial_g \lambda_k = 0$ and $\sum_k \lambda_k(\alpha_k + \beta_k) = 0$. The first condition is satisfied automatically due to normalization while the second one restricts $\alpha_k$ and $\beta_k$. Concurrently, condition (i) also places constraints on these quantities. Defining $|u\rangle = (A-\langle A \rangle)|\psi\rangle$, condition (i) translates to 
\[ {\rm Im}[\langle u | \partial_g \psi\rangle] = \sum_k \lambda_k(\lambda_k - \langle A \rangle)(\alpha_k + \beta_k) = 0. \] 

For Schmidt rank 2 which is of particular interest to us, using the gauge condition $\lambda_1 (\alpha_1 + \beta_1) + \lambda_2 (\alpha_2 + \beta_2) =0$, we have
\[ {\rm Im}[\langle u | \partial_g \psi\rangle] = \lambda_1\lambda_2(\lambda_1 - \lambda_2)[(\alpha_1 + \beta_1) - (\alpha_2 + \beta_2)] = 0. \] 
For $\lambda_1 \neq \lambda_2$ we then require $\alpha_1 + \beta_1 = \alpha_2 + \beta_2$ for satisfying condition (i). This means that the two Schmidt coefficients should not pick up a relative phase due to time evolution. Condition (ii) subsumes condition (i) in the case where Schmidt rank is 2. For rank greater than 2, there are $r-1$ relative phases and in this case condition (i) will constrain only one combination of the relative phases while condition (ii) will constrain all of them and so condition (ii) subsumes condition (i).

It is interesting to note that with the saturation conditions imposed, the Schmidt coefficients satisfy the continuous-time \emph{replicator equation} of evolutionary game theory, $\partial_g \lambda_k  = \lambda_k (2c \lambda_k- 2c \langle A \rangle)$ with $\langle A \rangle=\sum_j\lambda_j^2$ and fitness function $f_k = 2c \lambda_k$. This means each Schmidt weight grows or shrinks in proportion to how far its own value exceeds the mean (purity). The fitness is a gradient, $f_k=\partial_{\lambda_k}(c\sum_j\lambda_j^2)$, so the replicator equation is also the Riemannian gradient flow, with respect to the \emph{Fisher--Rao (Shahshahani) metric} $ds_{FR}^2=\sum_k d\lambda_k^2/\lambda_k$ on the probability simplex, of the purity $\Tr\rho_A^2$.

\end{document}